\documentstyle[12pt]{article}

\makeatletter
\def\vereq#1#2{\lower3pt\vbox{\baselineskip1.5pt \lineskip1.5pt
\ialign{$\m@th#1\hfill##\hfil$\crcr#2\crcr\sim\crcr}}}
\makeatother

\begin{document}

\begin{titlepage}
\begin{center}
\today     \hfill    LBNL-40346 \\
~{} \hfill UCB-PTH-97/21  \\
~{} \hfill hep-th/9705189\\

\vskip .1in

{\large \bf Renormalization Group Invariance of\\ Exact Results
in Supersymmetric Gauge 
Theories}\footnote{This work was supported in part by the U.S. 
Department of Energy under Contract DE-AC03-76SF00098, in part by the 
National Science Foundation under grant PHY-95-14797.  NAH was also 
supported by NERSC, and HM by Alfred P. Sloan Foundation.}

\vskip 0.1in

Nima Arkani-Hamed and Hitoshi Murayama

\vskip 0.05in

{\em Theoretical Physics Group\\
     Ernest Orlando Lawrence Berkeley National Laboratory\\
     University of California, Berkeley, California 94720}

\vskip 0.05in

and

\vskip 0.05in

{\em Department of Physics\\
     University of California, Berkeley, California 94720}

\end{center}


\begin{abstract}
  We clarify the notion of Wilsonian renormalization group (RG)
  invariance in supersymmetric gauge theories, which  states that the
  low-energy physics can be kept fixed when one changes the
  ultraviolet cutoff, provided appropriate changes are made to the
  bare coupling constants in the Lagrangian.  We first pose a puzzle
  on how a quantum modified constraint (such as $\mbox{Pf} (Q^i Q^j) =
  \Lambda^{2(N+1)}$ in SP($N$) theories with $N+1$ flavors) can be RG
  invariant, since the bare fields $Q^i$ receive wave function
  renormalization when one changes the ultraviolet cutoff, while we
  naively regard the scale $\Lambda$ as RG invariant.  The resolution
  is that $\Lambda$ is {\it not}\/ RG invariant {\it if}\/ one sticks
  to canonical normalization for the bare fields as is conventionally
  done in field theory.  We derive a formula for how $\Lambda$ must be
  changed when one changes the ultraviolet cutoff.  We then compare
  our formula to known exact results and show that their consistency
  requires the change in $\Lambda$ we have found.  Finally, we apply
  our result to models of supersymmetry breaking due to quantum
  modified constraints.  The RG invariance helps us to determine the
  effective potential along the classical flat directions found in 
  these theories. In   particular, the inverted hierarchy mechanism does 
  not occur in the original version of these models.
\end{abstract}

\end{titlepage}

\newpage
\section{Introduction}
\setcounter{equation}{0}
\setcounter{footnote}{0}

The last two years have seen remarkable progress in understanding the dynamics
of supersymmetric gauge theories (for a
review, see \cite{IS}).  It is now worthwhile to consider model building
implications of strong supersymmetric gauge dynamics, 
especially in the areas of composite models or dynamical supersymmetry
breaking.  Quantitative results are often required in many
phenomenological applications. For instance, the exact vacuum structure
and mass spectrum are needed for realistic models of dynamical supersymmetry
breaking. Similarly, the Yukawa couplings must be determined
in a realistic composite model. It is, therefore,
useful to have a closer look at the
quantitative results which follow from exactly solved supersymmetric 
gauge theories.

Actually, a detailed look at these exact results leads 
to some possible confusions.  For instance, the quantum
modified constraint in SP($N$) theories with $N+1$ flavors:
\begin{equation}
        \mbox{Pf} (Q^{i} Q^{j}) = \Lambda^{2(N+1)}
\end{equation}
appears inconsistent at the first sight.  The left-hand side involves
quantum fields which acquire wave function renormalization, while the
right-hand side appears renormalization group (RG) invariant.  

It is the purpose of this paper to clarify possible confusions
associated with the RG invariance of exact results.  RG analysis
always contains two steps.   The first step is naive dimensional
analysis which 
changes all dimensionful parameters by the same factor $e^{t}$; it in
particular changes the cutoff scale $M$ where the theory is defined to
$e^{t} M$.  The second is the readjustment of the bare parameters in the
Lagrangian as the cutoff scale is changed from $e^{t} M$
back to $M$, keeping the low energy physics fixed.  The first
part is of course trivial.  The second part requires care. 
We will show that the scale
$\Lambda$ actually changes when one changes the cutoff back to $M$,
and hence is {\it not}\/ RG invariant.  We will discuss in detail why
this is true and how an improved understanding helps to avoid possible
further confusions.  Although most of the essential ingredients in
this paper are already contained in the seminal work of Shifman and
Vainshtein \cite{SV}, we hope that our paper will help in clarifying
this subtle issue and in applying RG invariance to practical problems.
See also other analyses in Refs.~\cite{DS,FP}.

The main result of the paper is quite simple.  When one changes the
ultraviolet cutoff $M$ to $M'<M$ by integrating out a momentum slice,
and keeps the same form for the Lagrangian, {\it i.e.}\/ canonical
kinetic terms for the bare chiral superfields,\footnote{We stick to
  canonical normalization for chiral superfields simply because the
  first part of the RG analysis (naive scaling) preserves the
  normalization of the fields and therefore this choice makes the
  application of the RG analysis simpler.  When one would like to keep
  the holomorphy of the gauge coupling constant manifest, one needs to
  keep track of the wave function renormalization in a different
  manner.  The same results are obtained for physical quantities either
  way: see Section~\ref{sec:N=2} for an example.} one needs to replace the
 holomorphic gauge coupling constant\footnote{A holomorphic gauge
  coupling is defined by the coefficient of the $WW$ operator in the
  Lagrangian with $W_{\alpha} = \bar{D}^{2} e^{-2V_{h}} D_{\alpha}
  e^{2V_{h}}$.  We will explain how the holomorphic gauge coupling
  constant is related to the one in canonical normalization
  $W_{\alpha} = \bar{D}^{2} e^{-2g_c V_{c}} D_{\alpha} e^{2g_c V_{c}}$
  used in the perturbation theory in Section three.} as
\begin{equation}
  \frac{8\pi^{2}}{g_{h}^{2}} \rightarrow 
  \frac{8\pi^2}{g_h^{\prime 2}} = 
        \frac{8\pi^{2}}{g_{h}^{2}} + b_{0} \ln \frac{M}{M'} 
        - \sum_{i} T_{F}^{i} \ln Z_{i} (M', M)
        \label{eq:result}
\end{equation}
where $b_{0} = - 3 C_A + \sum_i T_F^i$ is the one-loop
$\beta$-function, $T_{F}^{i}$ is the $\beta$-function coefficient for a
chiral multiplet $T_{F}^{i} \delta^{ab}= \mbox{Tr}_{i} T^{a} T^{b}$,
and $Z_{i} (M',M)$ is the coefficient of the kinetic term for the
chiral multiplet $i$ when the modes between $M'$ and
$M$ are integrated out. 
Employing this formula, it is straightforward to check the
consistency of various results.  

The above formula implies that the dynamical scale which appears in
exact results:
\begin{equation}
  \label{eq:Lambda}
  \Lambda(M, 1/g_h^2)
\equiv M \exp(8\pi^{2}/g_{h}^{2}b_{0})
\end{equation}
is not RG invariant in theories with matter multiplets.  Under the
change of the cutoff and bare parameters, it changes to
\begin{equation}
  \label{eq:result2}
  \Lambda(M', 1/g_h^{\prime 2}) = \Lambda(M, 1/g_h^2) \prod_i
  Z_i(M,M')^{-T^i_F/b_0}. 
\end{equation}
This observation can solve possible confusions about the
RG invariance of exact results and effective potentials.  We point
out that a naive argument (regarding $\Lambda$ RG invariant) 
gives a qualitatively incorrect conclusion for the 
vacuum structure of a theory breaking supersymmetry dynamically. The
result is interesting for model building: the naive understanding allows the 
inverted hierarchy mechanism to be realized in these models, whereas the 
correct understanding shows that this is impossible.

The paper is organized as follows.  In the next section, we formulate
the Wilsonian renormalization program in the context of supersymmetric
gauge theories.  We derive the formula Eq.~(\ref{eq:result}) in this
section.  The same result is derived by perturbative calculations in
section three.  Section four describes various examples where our
formula guarantees the consistency of known results.  In section
five, we apply our improved understanding to a particular model where
one might naively expect the inverted hierarchy mechanism to work.
A careful application of our formalism demonstrates that this is not the
case.  We conclude in section six.

\section{Wilsonian Renormalization Group}
\label{sec:Wilsonian}
\setcounter{equation}{0}
\setcounter{footnote}{0}

In this section, we review the notion of Wilsonian Renormalization
Group and apply it to supersymmetric gauge theories.  Based on the
Shifman--Vainshtein \cite{SV} result that the renormalization of the
gauge kinetic term is exhausted at one-loop, and the anomalous
Jacobian of the path integral under the rescaling of the quantum
fields \cite{KS}, we determine the correct readjustment of
the bare couplings to derive Eq.~(\ref{eq:result}).

A field theory is normally defined by specifying the bare parameters
$\lambda^0_i$ and some cutoff scale $M$. All Green's functions
can then be calculated as functions of $\lambda^0_i$ and
$M$.  To work out Green's functions at energy scales much below the
cutoff $M$, it is convenient to ``integrate out'' physics between
$\mu$ and $M$ in a path integral and write down a new Lagrangian with
a cutoff $\mu$ which is close to the energy scale of the interest.
It is a non-trivial fact (related to the 
renormalizability of the theory) that one
does not need to specify infinite number of bare couplings for all
possible operators: those for relevant 
({\it i.e.}\/ dimension $\leq 4$) operators are enough to define the  
theory.  Therefore, one can define the RG flow of the finite number of
bare parameters as one changes the cutoff gradually by ``integrating
out'' modes.

Practically, we determine the bare couplings from experiments.  By
measuring the amplitudes corresponding to the relevant operators at
low energies $E$ (which we refer to loosely as $\lambda_i(E)$), we can
work backwards and determine what values of $\lambda^0_i$ are needed
to reproduce the measured $\lambda_i(E)$. If we work with a different
cutoff $M'$, but wish to reproduce the same observed values of the
$\lambda_i(E)$, a different set of bare parameters $\lambda^{0 \,
  \prime}_i$ must be chosen.  However, once this choice is made, the
predictions for all other low energy amplitudes are
identical
\footnote{
Actually, for the amplitudes for two theories with cutoffs $M,M'$ to be 
{\it exactly}\/ the same, an infinite number of higher dimension operators will 
have to be included in the definition of the theory with cutoff $M'$. However,
it is always possible to absorb the effects of higher dimension operators
into the relevant operators, up to calculable 
finite corrections suppressed by powers in $E/M$, $E/M'$.} 
whether we work with the theory based
on $(\lambda^0_i,M)$ or $ (\lambda^{0 \, \prime}_i,M')$.  

We should emphasize that none of our discussions depend on the precise
way in which the theory is cutoff at the scale $M$, nor the precise
way of integrating out modes.  The point is simply that it is possible
to change the bare couplings $\lambda_i^0$ with the cutoff $M$ while
keeping the low-energy physics fixed.  This is the formal definition
of the ``integrating out modes'' procedure.  The way in
which the $\lambda^0_i$ must change with $M$, while keeping the low energy
physics fixed, is encoded in Renormalization Group Equations (RGE's) for
the $\lambda^0_i$:
\begin{equation}
  M\frac {d}{d M} \lambda^0_i = \beta^0_i(\lambda^0) .
  \label{eq:WRGE}
\end{equation}

All of the usual results of RG analysis follow from the above
considerations. The procedure is always the same: for any quantity of
interest, first, one rescales all parameters (including the cutoff) by
naive dimensional analysis, then one changes the cutoff back to the
original one while simultaneously changing the bare couplings in
accordance with Eq. (\ref{eq:WRGE}).  As an example, consider the
1PI 4-point function $\Gamma^{4}(p_i;\lambda^0,M)$ for a
$\lambda^0 \phi^4/4!$ theory with cutoff $M$, and with all the
(Euclidean) momenta $|p_i| \sim \mu \ll M$.  If we just compute
$\Gamma^4$ in perturbation theory, we find
\begin{equation}
\Gamma^4(p_i;\lambda^0,M) = \lambda^0 - \frac{3}{16 \pi^2} \lambda^{0 \, 2} 
\mbox{ln} \frac{M}{\mu} + \cdots
\end{equation}
where $\cdots$ stands for higher order terms in perturbation theory and
non-logarithmic corrections which depend on $p_i$ and $\mu$.  For $\mu
\ll M$, the logarithm in the above becomes large and the 1-loop
term becomes comparable to the tree-level piece, making perturbation
theory unreliable. Let us now apply the procedure outlined above for
applying the Wilsonian RGE. First, we rescale everything by
dimensional analysis:
\begin{equation}
\Gamma^4(p_i;\lambda^0,M) = \Gamma^4(e^t p_i;\lambda^0,e^t M).
\end{equation}
Next, we use the Wilsonian RGE to bring the cutoff on the RHS of the above back from $e^t M$ to $M$ 
while changing $\lambda^0$ appropriately:
\begin{equation}
\Gamma^4(e^t p_i; \lambda^0,e^t M) = \Gamma^4(e^t p_i;\lambda(\lambda^0;t),M)
\end{equation}
where $\lambda(\lambda^0;t)$ is the solution of $(d/dt) \lambda = -
\beta^0(\lambda)$ with $\lambda(\lambda^0; 0)=\lambda^0$. We then have
\begin{eqnarray}
  \Gamma^4(p_i;\lambda^0,M) &=&\Gamma^4(e^t p_i;\lambda(\lambda^0;t),M) 
  \nonumber \\ 
  &=&
  \lambda(\lambda^0;t) - \frac{3}{16 \pi^2} \lambda(\lambda^0;t)^2 
  \mbox{ln} \frac{M}{\ e^t \mu} + \cdots
\end{eqnarray}
and if we choose $t$ so that $e^t \mu \sim M$, the logarithms on
the second line of the above are small and the perturbation expansion
is reliable. In particular we have the standard result
\begin{equation}
  \Gamma^4(p_i;\lambda^0,M) = 
  \lambda(\lambda^0,t \sim \mbox{ln} M/\mu) 
  + \mbox{small calculable corrections}.
\end{equation}

Let us now consider the Wilsonian RGE for supersymmetric gauge theories with
matter. With some cutoff $M$, the theory is specified by the bare
Lagrangian
\begin{equation}
  {\cal L}(M) = \frac{1}{4} \int d^2\theta \frac {1}{g_h^2} W^a W^a + h.c. 
  + \int d^4\theta \sum_i \phi_i^{\dagger} e^{2 V_i} \phi_i ,
\end{equation}
where $V_i = V^a T^a_i$, and $T^a_i$ are generators in the
representation of the chiral superfield $\phi_i$.  We are working with
the holomorphic normalization for the gauge coupling
\begin{equation}
\frac {1}{g_h^2} = \frac{1}{g^2} + i \frac{\theta}{8 \pi^2}.
\end{equation}
Actually, there is a hidden parameter in the above Lagrangian: the
coefficient of the matter field kinetic term $Z_i(M)$. However, we
have chosen to work with canonical normalization for the bare matter
field kinetic terms and we have set $Z_i(M)=1$.  There are two reasons
for taking canonical normalization:  (1) this is the conventional
choice in field theory, (2) it is easy to compare Lagrangians with
different cutoffs with fixed normalization of the bare fields, since
the naive dimensional analysis part of the RG analysis preserves the
normalization of the kinetic term.  Now with this choice of the
normalization, when we change the cutoff from $M$ to $M'$, how
should the bare parameters be changed to keep the low energy physics
fixed?  Shifman and Vainshtein argued that, to all orders of
perturbation theory, the couplings should be changed so that the
Lagrangian with cutoff $M'$ becomes
\begin{eqnarray}
  {\cal L}(M') &=& \frac{1}{4} \int d^2\theta 
  \left(\frac{1}{g^2_h} + 
    \frac{b_0}{8 \pi^2} \mbox{ln} \frac{M}{M'} \right)W^a W^a 
  + h.c. \nonumber \\
  & &
  + \int d^4\theta \sum_i Z_i(M,M') \phi_i^{\dagger} e^{2 V_i} \phi_i .
\end{eqnarray}
That is, the holomorphic coupling receives only 1-loop contributions.
However, the matter field kinetic terms do not remain canonical in
going from $M$ to $M'$. 

One can easily understand that the change of $1/g_h^2$ is exhausted at
1-loop in perturbation theory as long as the change is holomorphic.
This is because holomorphy and periodicity in $\theta$ demand that one
can expand the dependence in Fourier series of $\exp(- 8\pi^2/g_h^2)$,
\begin{equation}
  \frac{1}{g^2_h} + \sum_{n \geq 0} a_n\left(\frac{M}{M'}\right)
  \mbox{exp}\left(-n \frac{8 \pi^2}{g^2_h(M)}\right). 
\end{equation}
The sum is limited to the positive frequencies $n\geq 0$ to ensure
that the theory has a well-defined weak coupling limit $g^2_h
\rightarrow 0$.  The terms with $n>0$ can never
arise in perturbation theory, and we drop them. 
The function $a_0(M/M')$ must
satisfy the consistency condition $a_0(M/M') + a_0(M'/M'') =
a_0(M/M'')$, and hence it must be a logarithm.  This proves the
one-loop law of the change in holomorphic gauge coupling constant.

The point is, however, that the change in $1/g^2_h$ is holomorphic
only when the normalization for the matter field kinetic terms (which
is manifestly non-holomorphic, being only a function of $g$) is
allowed to change from 1 to $Z(M,M')$.

In order to go back to canonical normalization for the matter fields,
one simply redefines $\phi = Z(M,M')^{-1/2} \phi'$.  However, the path
integral measure $D\phi$ is not invariant under this change,
$D(Z(M,M')^{-1/2} \phi') \neq D \phi'$; there is an anomalous Jacobian
\cite{KS}.  In our case, $Z(M,M')$ is positive and real, but it is
sensible to look at $D(Z^{-1/2} \phi')$ for a general complex number
$Z$ since $\phi'$ is a chiral superfield. When $Z=e^{i \alpha}$ is a
pure phase, the field redefinition is a chiral rotation on the
fermionic component of $\phi'$ and the Jacobian is the one associated
with the chiral anomaly.  This Jacobian is exactly known \cite{KS} and
is cutoff independent:
\begin{eqnarray}
  \lefteqn{
    D(e^{-i \alpha/2} \phi') D(e^{+i \alpha/2} \phi'^{\dagger}) }
  \nonumber \\
  &=& D\phi' D\phi^{\prime\dagger}
  \exp\left(\frac{1}{4} \int d^4 y \int d^2 \theta
  \frac{T_F(\phi)}{8 \pi^2} \ln(e^{i \alpha}) W^a W^a + h.c. \right) .
\label{eq:purephase}
\end{eqnarray}
In the case where $Z$ is a general complex number, the Jacobian will
in general have $F$ terms and $D$ terms (such as $Re(\ln Z) W^* W^* W
W$).  However, since $R$ symmetry is at least good in perturbation
theory, the $F$ terms can only contain $W^a W^a$, and its coefficient is
the same as in Eq.~(\ref{eq:purephase}) with $\ln e^{i \alpha}$ replaced
by $\ln Z$. The $D$ terms are all higher dimensional operators
suppressed by powers of the cutoff and can be neglected.\footnote{In a
  general non-supersymmetric theory, it is not possible to simply
  throw away higher dimension operators suppressed by the cutoff,
  since loops with these operators may contain power divergences which
  negate the cutoff suppression; what {\it can} be done is to set the
  operators to zero with an appropriate modification of the
  renormalizable couplings. However, in supersymmetric theories, the
  non-renormalization theorem makes it impossible for the higher
  dimensional $D$ terms to ever contribute to the coefficient of $W^a W^a$
  which is an $F$ term, and so the higher dimensional $D$ terms really
  can be dropped \cite{AM2}}

Therefore, if we wish to keep canonical normalization for the matter
fields in changing the cutoff from $M$ to $M'$, the Lagrangian at
cutoff $M'$ must be given by
\begin{equation}
  {\cal L}'(M') = \frac{1}{4}\int d^2\theta \frac{1}{g'^2_h} W^a W^a + h.c.+ \int
  d^4\theta \sum_i \phi_i^{\dagger} e^{2 V_i} \phi_i 
\end{equation}
where 
\begin{equation}
  \frac{1}{g'^2_h} = \frac {1}{g_h^2} + 
    \frac{b_0}{8 \pi^2} \ln \frac{M}{M'}
    - \sum_i \frac{T_F(\phi^i)}{8 \pi^2} \ln Z_i(M,M').
\end{equation}
We can rephrase the above results in terms of the scale $\Lambda(M,
1/g_h^2)$ (see Eq.~(\ref{eq:Lambda})).  If we change the cutoff from
$M$ to $M'$, and always work with canonical normalization for the
matter fields, we have
\begin{equation}
  \Lambda(M, 1/g_h^2) \rightarrow \Lambda(M', 1/g'^2_h)
  =\Lambda(M, 1/g_h^2) \prod_i Z_i(M,M')^{-T^i_F/b_0}. 
\end{equation}  

So far we have considered the case with zero superpotential, but the
extension to the general case is obvious. For instance, suppose we add
a superpotential term of the form $\int d^2 \theta W=\int d^2 \theta
\lambda^{i j k} \phi_i \phi_j \phi_k$.  Then by the
non-renormalization theorem, $\lambda^{i j k}$ stays the same if we
allow non-canonical kinetic terms  We, however, insist on working with
canonical kinetic terms, and we must have $\lambda'^{i j k} =
Z_i(M,M')^{-1/2} Z_j (M,M')^{-1/2} Z_k(M,M') ^{-1/2} \lambda^{i j k}$.

\section{Perturbative Derivation}
\setcounter{equation}{0}
\setcounter{footnote}{0}

In this section, we rederive the result obtained in the previous
section by perturbative calculations.  We first review how one can
relate perturbative results to the exact results, and then discuss how
we change the bare parameters as we change the ultraviolet cutoff.
The final result is the same as Eq.~(\ref{eq:result}).\footnote{Note
  that the analysis in this section is not independent from the one in
  the previous section; it is simply a reanalysis in a different
  language.  The one-loop exhaustion of the renormalization of $W^a W^a$
  used in the previous section and NSVZ $\beta$-function  used in this
  section are closely related \cite{AM2}.}

Comparison of the perturbative results to the exact results is a
somewhat confusing issue.  The so-called anomaly puzzle is one famous
example of such a confusion.  In supersymmetric theories, the
U(1)$_{R}$ current belongs to the same supermultiplet as the trace of
the energy-momentum tensor, and hence the chiral anomaly and the trace
anomaly are related.  On the other hand, the chiral anomaly is
exhausted at one-loop (Adler--Bardeen theorem) while the trace anomaly
is not in $N=1$ theories.  Shifman and Vainshtein made a breakthrough
on this question by discriminating two definitions of coupling
constants: ``canonical'' and ``holomorphic''.\footnote{We find the
  terminology by Shifman and Vainshtein rather confusing.  In our
  understanding, what they call ``1PI'' coupling constant is not what
  appears in 1PI effective actions; they are still coupling constants
  in Wilsonian effective action.  The only difference between them is
  that one employs canonical normalization for gauge field kinetic
  term in ``1PI'' couplings while holomorphic normalization in
  ``Wilsonian'' couplings.  We will rather refer to them as
  ``canonical'' and ``holomorphic'' gauge coupling constants in this
  paper.  We will discuss more on this issue in our forthcoming paper
  \cite{AM2}.} The holomorphic gauge coupling $g_{h}$ runs only at
one-loop, while the canonical gauge coupling $g_{c}$ has higher order
$\beta$-functions.  There is a simple relation between them, the 
Shifman--Vainshtein formula,
\begin{equation}
        \frac{8\pi^{2}}{g_{h}^{2}} = \frac{8\pi^{2}}{g_{c}^{2}}
                + C_{A} \ln g_{c}^{2},
        \label{eq:SV}
\end{equation}
where $f^{acd} f^{bcd} = C_{A} \delta^{ab}$.  The difference $C_{A}
\ln g^{2}$ appears due to an anomalous Jacobian in the path integral
when one rescales the vector multiplet $V_{h}$ which appears in the
field strength $W_{\alpha} = \bar{D}^{2} e^{-2V_{h}} D_{\alpha}
e^{2V_{h}}$ to the one in canonical normalization $V_{h}=g_{c}V_{c}$.
The Lagrangian written in terms of $V_{h}$ does not need the gauge
coupling constant in the exponent, and hence does not need to
separate the $\theta$ angle from the gauge coupling constant.  This
normalization of the vector multiplet therefore keeps holomorphicity
of the gauge coupling constant manifest (holomorphic normalization)
while the canonical one requires an explicit dependence on the gauge
coupling constant in the exponent.

There still remains the question how the canonical gauge coupling
constant $g_c$ in the Wilsonian action is related to the perturbative
definitions of the running coupling constant in popular schemes such
as $\overline{\rm DR}$.  We are not aware of a complete answer to this
question,\footnote{At least in some models, one can define a
  regularized Wilsonian action of the theory and compare the
  canonical gauge coupling constant in the Wilsonian action to the
  perturbative definition \cite{AM2}.  } even though one can work out
the relation between the two coupling constants at each order in
perturbation theory \cite{JJN}.  

There is a known ``exact'' $\beta$-function in supersymmetric gauge
theories by Novikov--Shifman--Vainshtein--Zakharov (NSVZ) \cite{NSVZ}.
Our understanding is that this exact $\beta$-function applies to the
canonical gauge coupling constant in a Wilsonian action and is hence
appropriate for our analysis \cite{AM2}.  Therefore we employ the NSVZ
$\beta$-function for our perturbative analysis to determine the
necessary change of the bare parameters to keep the low-energy physics
fixed as we change the ultraviolet cutoff.  The exact NSVZ
$\beta$-function is given by
\begin{equation}
  \mu \frac{d g^2}{d\mu} = 
        \beta = - \frac{g^4}{8\pi^2}
                \frac{3 C_{A} - \sum_{i} T_{F}^{i}(1-\gamma_{i})}
                        {1-C_{A} g^2/8\pi^2} ,
        \label{eq:NSVZ}
\end{equation}
with $\gamma_i = (\mu d/d\mu) \ln Z_i(\mu, M)$.  Of course the
$\beta$-function for the gauge coupling constant is the same up to
two-loop order in any schemes.  From the strictly perturbative point
of view, one can regard our analysis as a two-loop analysis in,
say, $\overline{\rm DR}$.  The RGE can then be integrated with the
NSVZ $\beta$-function, and one finds that
\begin{eqnarray}
\lefteqn{
        \left( \frac{8\pi^{2}}{g_c^{2}(\mu)} + C_{A} \ln g_c^{2}(\mu) 
        + \sum_{i} T_{F}^{i} \ln Z_{i}(\mu,M) \right) }\nonumber \\
& &     = \left( \frac{8\pi^{2}}{g_c^{2}(M)} + C_{A} \ln g_c^{2}(M) \right)
        + b_{0} \ln \frac{M}{\mu} .
        \label{eq:running}
\end{eqnarray}
The combination in the bracket runs only at one-loop, and the 
wavefunction renormalization factors are by definition unity at 
the cutoff scale, $Z_{i}(M,M) = 1$.  The $\beta$-function coefficient 
is given by $b_{0} = - 3 C_{A} + \sum_{i} T_{F}^{i}$.

Now the strategy is to change the bare parameters $M$ and 
$g_c^{2}(M)$ while keeping the low-energy physics ($g_c^{2}(\mu)$)
fixed.  Naively, the change required appears to come from $b_{0} 
\ln (M/\mu)$ in the right-hand side, and the change
\begin{equation}
        \left( \frac{8\pi^{2}}{g_c^{2}(M')} + C_{A} \ln g^{2}_c(M') \right)
        = \left( \frac{8\pi^{2}}{g^{2}_c(M)} + C_{A} \ln g^{2}_c(M) \right)
        + b_{0} \ln \frac{M}{M'}
\end{equation}
might appear to be enough.  However, this is not correct, because 
the wave function renormalization factor $Z_{i}(\mu,M)$ in the 
left-hand side also depends on $M$ implicitly due to the boundary 
condition $Z_{i}(M,M)=1$.

The trick is that the wave function renormalization is 
multiplicative:
\begin{equation}
        Z_{i} (\mu, M) = Z_{i} (\mu, M') Z_{i} (M', M).
\end{equation}
Then Eq.~(\ref{eq:running}) can be rewritten as
\begin{eqnarray}
\lefteqn{
        \left( \frac{8\pi^{2}}{g^{2}_c(\mu)} + C_{A} \ln g^{2}_c(\mu) 
        + \sum_{i} T_{F}^{i} \ln Z_{i}(\mu,M') \right) } \nonumber \\
&=&     \left( \frac{8\pi^{2}}{g^{2}_c(M)} + C_{A} \ln g^{2}_c(M) \right)
        + b_{0} \left(\ln \frac{M'}{\mu} +\ln \frac{M}{M'} \right)
        - \sum_{i} T_{F}^{i} \ln Z_{i}(M',M). \nonumber \\
        \label{eq:running2}
\end{eqnarray}
It is now clear that the correct change of the bare parameters 
is
\begin{eqnarray}
\lefteqn{
        \left( \frac{8\pi^{2}}{g^{2}_c(M')} + C_{A} \ln g^{2}_c(M') \right)
        } \nonumber \\
& &     = \left( \frac{8\pi^{2}}{g^{2}_c(M)} + C_{A} \ln g^{2}_c(M) \right)
        + b_{0} \ln \frac{M}{M'} - \sum_{i} T_{F}^{i} \ln Z_{i}(M',M) ,
\end{eqnarray}
which keeps the low-energy physics ($g_c^2(\mu)$) fixed.

The final step is to rewrite the above relation in terms of the 
holomorphic gauge coupling $g_{h}$ using the Shifman--Vainshtein 
formula Eq.~(\ref{eq:SV}),
\begin{equation}
        \frac{8\pi^{2}}{g_{h}^{2}}(M') =
        \frac{8\pi^{2}}{g_{h}^{2}}(M) + b_{0} \ln \frac{M}{M'} 
        - \sum_{i} T_{F}^{i} \ln Z_{i} (M', M)
\end{equation}
This is indeed the same relation as obtained in the previous 
section.

If a chiral superfield has a coupling $\lambda^{ijk}$ in the
superpotential, it is renormalized only due  
to wave function renormalization because of the non-re\-norm\-ali\-za\-tion 
theorem.  The low-energy coupling is given by
\begin{equation}
        \lambda^{ijk} (\mu) = \lambda^{ijk} Z_i^{-1/2}(\mu, M)
        Z_j^{-1/2}(\mu, M) Z_k^{-1/2}(\mu, M) .
\end{equation}
By using the multiplicative property of the wave function 
renormalization again, the change of the bare parameter is
\begin{equation}
        \lambda'^{ijk} = \lambda^{ijk} Z_i^{-1/2}(M', M)
        Z_j^{-1/2}(M', M) Z_k^{-1/2}(M', M)
        \label{eq:mchange}
\end{equation}
to keep $\lambda^{ijk}(\mu)$ fixed when one changes the cutoff.  

\section{Examples}
\setcounter{equation}{0}
\setcounter{footnote}{0}

In this section, we apply our result
Eqs.~(\ref{eq:result},\ref{eq:result2}) to many examples.  The RG
invariance is checked usually with two steps, (1) naive dimensional
analysis, and (2) the change of cutoff parameters.  For simplicity of
the presentation, we do not discuss the first part since it is rather
trivial.  The non-trivial part of the analysis is the correct
application of the change of bare parameters as derived in previous
sections.

\subsection{Quantum Modified Moduli Space (I)}

In SP($N$) theories with $N+1$ flavors, Intriligator and Pouliot found 
the quantum modified constraint \cite{SP(N)}
\begin{equation}
        \mbox{Pf}(Q^{i} Q^{j}) = \Lambda^{2(N+1)} .
\end{equation}
Dine and Shirman \cite{DS} correctly emphasized that the fields in the
left-hand side are bare fields in a Wilsonian action with an
ultraviolet cutoff $M$.  The Lagrangian of the model is simply
\begin{equation}
        {\cal L} = \int d^{2} \theta \, \frac{1}{4 g_{h}^{2}} W^a W^a + h.c.
                + \int d^{4} \theta \, Q^{i\dagger} e^{2V} Q^{i}
\end{equation}
in terms of bare fields.

As explained in Section~\ref{sec:Wilsonian}, a Wilsonian RG allows the
change of ultraviolet cutoff while keeping the low-energy physics
fixed by appropriately changing the bare coupling constants in the
theory.  With the same Lagrangian given at a different cutoff $M'$, a
coupling constant $g'_{h}$, and bare fields $Q^{\prime i}$, we must
find
\begin{equation}
        \mbox{Pf}(Q^{\prime i} Q^{\prime j}) = \Lambda^{2(N+1)} .
                \label{eq:inconsistent}
\end{equation}
if $\Lambda$ were a RG invariant quantity.  Note 
that we need to keep the form of the Lagrangian the same no matter 
how we change the cutoff; therefore the fields $Q'$ must have 
canonical kinetic terms as $Q$ do.

The relation between the bare fields in two different Lagrangians, $Q$ 
and $Q'$ can be calculated.  When one integrates out modes between 
$M'$ and $M$, the original bare fields $Q$ acquire corrections to the
kinetic terms by a factor $Z_{Q} (M', M)$.  The bare fields $Q'$ have 
canonical normalization in the Lagrangian with the cutoff $M'$, and 
hence they are related by
\begin{equation}
        Q' = Z_{Q}^{1/2} (M', M) Q .
\end{equation}
Therefore, the left-hand sides of the constraint equations are related 
by
\begin{equation}
        \mbox{Pf} (Q'^{i} Q'^{j}) = Z_{Q}^{N+1}(M', M)
                \mbox{Pf} (Q^{i} Q^{j}),
\end{equation}
and hence the right-handed sides must also differ by 
$Z_{Q}^{N+1}$. Then Eq.~(\ref{eq:inconsistent}) is inconsistent.

Our result (\ref{eq:result2}) says that the dynamical scale of the
theory with cutoff $M'$ is related to the original one by
\begin{equation}
  \Lambda' = \Lambda \left[ Z_Q^{- T_F/b_0}(M', M)\right]^{2 N_f}
  = \Lambda Z_Q^{1/2}(M', M)
\end{equation}
with $T_{F} = 1/2$ and $b_0 = -2(N+1)$.  Now it is easy to see that
the quantum modified constraint holds between the primed fields and
the primed dynamical scale:
\begin{equation}
        \mbox{Pf}(Q^{\prime i} Q^{\prime j}) = \Lambda^{\prime 2(N+1)} .
\end{equation}
This is a consistency check that the quantum modified constraint 
is RG invariant.

\subsection{Quantum Modified Moduli Space (II)}

It is amusing to see how the quantum modified constraints are 
RG invariant in more complicated cases.  Let us 
look at SU($2k+1$) models with one anti-symmetric tensor $A$, three 
fundamentals $Q^{a}$ ($a=1,2,3$) and $2k$ anti-fundamentals 
$\tilde{Q}_{i}$ ($i=1, \ldots, 2k$) \cite{PT}.  The moduli space can 
be described by the gauge invariant polynomials
\begin{eqnarray}
M_i^a ~&=&~ \tilde{Q}^\alpha_i~ Q_\alpha^a \nonumber \\
X_{ij}  ~&=&~A_{\alpha \beta} ~\tilde{Q}^\alpha_i ~\tilde{Q}^\beta_j \nonumber \\
Y^a ~&=&~ Q_{\alpha_{2 k + 1}}^a ~\epsilon^{\alpha_1 ...\alpha_{2 k + 1}}
{}~A_{\alpha_1 \alpha_2} ... A_{\alpha_{2 k - 1} \alpha_{2 k}}  \\
Z ~&=&~ \epsilon^{\alpha_1 ...\alpha_{2 k + 1} }~
A_{\alpha_1 \alpha_2} ... A_{\alpha_{2 k - 3} \alpha_{2 k -2 }}
{}~Q^a_{\alpha_{2 k - 1}}~ Q^b_{\alpha_{2 k}} ~ Q^c_{\alpha_{2 k + 1}}~
\epsilon_{a b c} \nonumber ~,
\label{eq:oddinvts}
\end{eqnarray}
and the quantum modified constraint
\begin{equation}
\label{eq:oddquantumconstraint}
Y \cdot M^2 \cdot X^{k - 1} ~-~\frac{k}{3}~ Z ~
{\rm Pf} X ~=~\Lambda^{4 k + 2}~.
\end{equation}
By following Eq.~(\ref{eq:result2}), we find
\begin{equation}
        \Lambda^{\prime -b_{0}} = \Lambda^{-b_{0}}
                Z_{A}^{k-1/2} Z_{Q}^{3/2} Z_{\tilde{Q}}^{k}
\end{equation}
with $b_{0} = -(2k+1)$.  The quantum modified constraint is indeed RG
invariant as we change the cutoff from $M$ to $M'$, replacing all
fields by primed fields (with canonical kinetic terms) {\it and}\/ the
dynamical scale $\Lambda$ by $\Lambda'$.

\subsection{Matching Equations (I)}

When there is a massive chiral superfield, the gauge coupling 
constants in a theory with a massive field (high-energy theory) and 
the other theory where the massive field is integrated out (low-energy 
theory) are related by matching equations.  For SU($N$) gauge group 
with a single massive vector-like pair in the fundamental ($Q$) and 
anti-fundamental ($\tilde{Q}$) representations, the holomorphic 
coupling constants in high-energy ($g^{2}_{h,HE}$) and low-energy 
($g^{2}_{h,LE}$) theories are related by
\begin{equation}
        \frac{8\pi^{2}}{g^{2}_{h,LE}}
        = \frac{8\pi^{2}}{g^{2}_{h,HE}} + \ln \frac{M}{m}
\end{equation}
where $m$ is the bare mass of the field.  This form can be completely 
fixed (up to a possible constant) by the holomorphy in 
$8\pi^{2}/g_{h}^{2}$ and $m$, and the anomaly under the chiral U(1) 
rotation of the matter fields.  We drop the
possible constant in the following equations, and it can be easily
recovered if necessary.

Under the change of the cutoff, we rewrite the left-hand side as
\begin{equation}
        \frac{8\pi^{2}}{g^{\prime 2}_{h,LE}}
        = \frac{8\pi^{2}}{g^{2}_{h,LE}} + b_{0,LE} \ln \frac{M}{M'}
        - \sum_{i\neq Q,\tilde{Q}} T_{F}^{i} \ln Z_{i}(M', M)
\end{equation}
where the sum does not include the massive field $Q$, $\tilde{Q}$ 
which are integrated out in the low-energy theory.  The coupling in the 
high-energy theory is also rewritten as
\begin{equation}
        \frac{8\pi^{2}}{g^{\prime 2}_{h,HE}}
        = \frac{8\pi^{2}}{g^{2}_{h,HE}} + b_{0,HE} \ln \frac{M}{M'}
        - \sum_{i} T_{F}^{i} \ln Z_{i}(M', M) ,
\end{equation}
but here the sum includes the massive field.  The $\beta$-functions are 
related as $b_{0,HE} = b_{0,LE} + 1$.  Now the matching equation reads 
as
\begin{eqnarray}
        \frac{8\pi^{2}}{g^{\prime 2}_{h,LE}}
        &=& \frac{8\pi^{2}}{g^{\prime 2}_{h,HE}} - \ln \frac{M}{M'}
        + \frac{1}{2} (\ln Z_{Q}(M', M) + \ln Z_{\tilde{Q}}(M', M))
        + \ln \frac{M}{m} \nonumber \\
        &=& \frac{8\pi^{2}}{g^{\prime 2}_{h,HE}} 
        + \ln \frac{M'}{m'}
\end{eqnarray}
with $m' = Z^{-1/2}_{Q}(M', M) Z^{-1/2}_{\tilde{Q}}(M', M) m$.  
Therefore, the matching equation takes the same form with the new 
cutoff and bare parameters.

One can also check the consistency with the perturbative calculations
on matching of canonical gauge coupling constants.  For instance in
$\overline{\rm DR}$ scheme, the one-loop matching
equation\footnote{Recall that one-loop matching is required when one
  employs two-loop RGE.  We are not aware of $\overline{\rm DR}$
  calculations of two-loop matching which can tell us whether $m_{r}$
  must be the on-shell mass or $\overline{\rm DR}$ mass where the
  latter is more likely.} is simply $g^{2}_{c,LE}(m_{r}) =
g^{2}_{c,HE}(m_{r})$, where $m_{r}$ is the renormalized mass of the
chiral multiplet.  By using the Shifman--Vainshtein relation
(\ref{eq:SV}) between the canonical and holomorphic gauge couplings
and NSVZ exact $\beta$-function (using the integrated form
Eq.~(\ref{eq:running})), the matching condition between the canonical
gauge couplings for high-energy and low-energy theories can be
obtained as
\begin{eqnarray}
        \frac{8\pi^{2}}{g^{2}_{c,LE}(\mu)}
        &=& \frac{8\pi^{2}}{g^{2}_{c,HE}(\mu)} + \ln \frac{\mu}{m}
        + \frac{1}{2} (\ln Z_{Q}(\mu, M) + \ln Z_{\tilde{Q}}(\mu, M))
        \nonumber \\
        &=& \frac{8\pi^{2}}{g^{2}_{c,HE}(\mu)} + \ln \frac{\mu}{m(\mu)}\, .
\end{eqnarray}
Therefore the gauge coupling constants can be matched at the 
renormalized mass of the heavy field $\mu = m(\mu) = m Z_{Q}^{-1/2}(\mu, 
M) Z^{-1/2}_{\tilde{Q}}(\mu, M)$ as expected.

\subsection{Matching Equations (II)}

When a chiral superfield acquires an expectation value and the Higgs 
mechanisms occurs, the gauge coupling constants in a theory with the 
full gauge group (high-energy theory) and the other theory only with 
unbroken gauge group (low-energy theory) are related by matching 
equations.  For SU($N$) gauge group with an expectation value of a 
single vector-like pair in the fundamental and anti-fundamental 
representations $Q$ and $\tilde{Q}$, they can acquire an expectation 
value along the $D$-flat direction $Q=\tilde{Q}$ and the gauge group 
breaks down to SU($N-1$).  The holomorphic coupling constants in 
high-energy ($g^{2}_{h,HE}$) and low-energy ($g^{2}_{h,LE}$) theories 
are related by
\begin{equation}
        \frac{8\pi^{2}}{g^{2}_{h,LE}}
        = \frac{8\pi^{2}}{g^{2}_{h,HE}} - \ln \frac{M^{2}}{\tilde{Q}Q}
\end{equation}
where $\tilde{Q}$, $Q$ are the bare fields.  This form can be
completely fixed (up to a possible constant) by the holomorphy in
$8\pi^{2}/g_{h}^{2}$ and $\tilde{Q}$, $Q$, non-anomalous vector U(1)
symmetry, and and the anomaly under the chiral U(1) rotation of the
matter fields.  We drop the possible constant in the following
equations, and it can be easily recovered if necessary.

Under the change of the cutoff, we rewrite the left-hand side as
\begin{equation}
        \frac{8\pi^{2}}{g^{\prime 2}_{h,LE}}
        = \frac{8\pi^{2}}{g^{2}_{h,LE}} + b_{0,LE} \ln \frac{M}{M'}
        - \sum_{i\neq Q,\tilde{Q}} T_{F}^{i} \ln Z_{i}(M', M)
\end{equation}
where the sum does not include the massive field $Q$, $\tilde{Q}$ 
which are integrated out in the low-energy theory.  The coupling in the 
high-energy theory is also rewritten as
\begin{equation}
        \frac{8\pi^{2}}{g^{\prime 2}_{h,HE}}
        = \frac{8\pi^{2}}{g^{2}_{h,HE}} + b_{0,HE} \ln \frac{M}{M'}
        - \sum_{i} T_{F}^{i} \ln Z_{i}(M', M) ,
\end{equation}
but here the sum includes the massive field.  The $\beta$-functions
are related as $b_{0,HE} = b_{0,LE} -2$.  Now the matching equation
reads as
\begin{eqnarray}
        \frac{8\pi^{2}}{g^{\prime 2}_{h,LE}}
        &=& \frac{8\pi^{2}}{g^{\prime 2}_{h,HE}} + 2 \ln \frac{M}{M'}
        + \frac{1}{2} (\ln Z_{Q}(M', M) + \ln Z_{\tilde{Q}}(M', M))
        - \ln \frac{M^{2}}{\tilde{Q}Q} \nonumber \\
        &=& \frac{8\pi^{2}}{g^{\prime 2}_{h,HE}} 
        - \ln \frac{M'^{2}}{\tilde{Q}'Q'}
\end{eqnarray}
where the primed fields are defined by
\begin{equation}
        \tilde{Q}' = Z_{\tilde{Q}}^{1/2}(M',M) \tilde{Q},
        \hspace{1cm}
        Q' = Z_{Q}^{1/2}(M',M) Q.
\end{equation}
Therefore, the matching equation takes the same form with the new 
cutoff and bare fields.  

One can also check the consistency with the perturbative calculations
on matching of canonical gauge coupling constants.  For instance in
$\overline{\rm DR}$ scheme, the one-loop matching
equation\footnote{Here again we are not aware of $\overline{\rm DR}$
  calculations of two-loop matching which can tell us whether $m_V$
  must be the on-shell mass or $\overline{\rm DR}$ mass.} is simply
$g^{2}_{c,LE}(m_V) = g^{2}_{c,HE}(m_V)$, where $m_V$ is the
renormalized mass of the heavy gauge multiplet.  By using the
Shifman--Vainshtein relation (\ref{eq:SV}) between the canonical and
holomorphic gauge couplings and NSVZ exact $\beta$-function (using the
integrated form Eq.~(\ref{eq:running})), the matching condition
between the canonical gauge couplings for high-energy and low-energy
theories can be obtained as
\begin{eqnarray}
\lefteqn{
        \frac{8\pi^{2}}{g^{2}_{c,LE}(\mu)}
        + (N-1) \ln g^{2}_{c,LE}(\mu)
        } \nonumber \\
        &&= \frac{8\pi^{2}}{g^{2}_{c,HE}(\mu)} 
        + N \ln g^{2}_{c,HE}(\mu)
        - \ln \frac{\mu^{2}}{\tilde{Q}Q}
        + \frac{1}{2} (\ln Z_{Q}(\mu, M) + \ln Z_{\tilde{Q}}(\mu, M)),
        \nonumber \\
\end{eqnarray}
and hence
\begin{eqnarray}
        \frac{8\pi^{2}}{g^{2}_{c,LE}(\mu)}
        = \frac{8\pi^{2}}{g^{2}_{c,HE}(\mu)} 
        + \left(N-\frac{1}{2}\right) 
        \ln \frac{g^{2}_{c,HE}(\mu)}{g^{2}_{c,LE}(\mu)}
        + \ln \frac{m_{V}^{2}(\mu)}{\mu^{2}} .
\end{eqnarray}
Here, the renormalized gauge boson mass $m_{V}$ is defined by
\begin{equation}
        m_{V}^{2}(\mu) \equiv g_{c,LE}(\mu) g_{c,HE}(\mu) Z^{1/2}_{Q}(\mu, 
        M) Z^{1/2}_{\tilde{Q}}(\mu, M) \tilde{Q}Q .
\end{equation}
The matching is particularly simple: $g^{2}_{c,LE}(\mu) = 
g^{2}_{c,HE}(\mu)$ at the renormalized gauge boson mass $\mu = 
m_{V}(m_{V})$ as expected.

\subsection{Affleck--Dine--Seiberg superpotential}

In SU($N$) gauge theories with $N_{f} < N$, a non-perturbative 
superpotential is generated,
\begin{equation}
        W = \frac{\Lambda^{(3N-N_{f})/(N-N_{f})}}
                {(\mbox{det}\tilde{Q}^{i} Q^{j})^{1/(N-N_{f})}} \, .
\end{equation}
Under the change of the cutoff, we find
\begin{equation}
        \Lambda^{\prime (3N-N_{f})} = \Lambda^{(3N-N_{f})}
                Z_{Q}^{N_{f}/2} Z_{\tilde{Q}}^{N_{f}/2},
\end{equation}
and the superpotential becomes
\begin{equation}
        W' = \frac{\Lambda^{\prime (3N-N_{f})/(N-N_{f})}}
                {(\mbox{det}\tilde{Q}^{\prime i} Q^{\prime j})^{1/(N-N_{f})}}
        = \frac{\Lambda^{(3N-N_{f})/(N-N_{f})}
                (Z_{Q}Z_{\tilde{Q}})^{N_{f}/2(N-N_{f})}}
                {(\mbox{det}\tilde{Q}^{i} Q^{j})^{1/(N-N_{f})}
                (Z_{Q}Z_{\tilde{Q}})^{N_{f}/2(N-N_{f})}}
        = W .
\end{equation}
The Affleck--Dine--Seiberg superpotential is RG 
invariant.

\subsection{Gaugino Condensate}

When all chiral superfields are massive, they can be integrated out 
from the theory and the low-energy pure Yang--Mills theory develops a 
gaugino condensate.  After matching the gauge coupling constant at the 
threshold, the size of the gaugino condensate is a function of the 
bare mass of the chiral superfields and the bare gauge coupling 
constant.  

If there are $N_{f}$ chiral superfields with the same mass $m$ coupled 
to SU($N$) gauge group, the size of the gaugino condensate can be 
calculated as
\begin{equation}
        \langle \lambda \lambda \rangle = m^{N_{f}/N} \Lambda^{3-N_{f}/N}
\end{equation}
using holomorphy and U(1)$_R$ symmetry up to an overall constant.
Under the change of the cutoff and bare parameters, the scale
$\Lambda$ changes to
\begin{equation}
        \Lambda^{\prime -b_{0}} = \Lambda^{-b_{0}} Z_{Q}^{N_{f}/2} 
        Z_{\tilde{Q}}^{N_{f}/2} .
\end{equation}
Here, $b_{0} = -(3 N - N_{f})$.  The corresponding change of the bare 
mass parameter is
\begin{equation}
        m' = m Z_{Q}^{-1/2} Z_{\tilde{Q}}^{-1/2} .
\end{equation}
It is easy to see that the gaugino condensate is invariant under these 
changes.

\subsection{$N=2$ theories}
\label{sec:N=2}

An application of our formalism to $N=2$ theories requires care
because of a difference in conventions.  Take $N=2$ supersymmetric QCD
with $N_f$ hypermultiplets in fundamental representation.  In $N=1$
language, the particle content of the theory is the vector multiplet
$V$, a chiral multiplet in the adjoint representation $\phi$, $N_f$
chiral multiplets $Q_i$ and $\tilde{Q}_i$ ($i=1, \ldots, N_f$) in
fundamental and anti-fundamental representations, respectively.  The
Lagrangian in the conventional normalization of fields in the $N=2$
context is
\begin{eqnarray}
  {\cal L} &=& \int d^4 \theta \left( Re\left(\frac{1}{g_2^2}\right)
    2\mbox{Tr} \phi_2^\dagger e^{2V} \phi_2 e^{-2V} + Q_i^\dagger e^{2V} Q_i
    + \tilde{Q}_i^\dagger e^{-2 V^T} \tilde{Q}_i \right)
  \nonumber \\
  & & + \int d^2 \theta \left( \frac{1}{4 g_2^2} W^a W^a + 
    \sqrt{2} \tilde{Q}_i \phi Q_i \right) + h.c.
\end{eqnarray}
Note that the normalization of the $\phi$ kinetic term is not
canonical.  Here we use the notation $1/g_2^2$ to refer to the gauge
coupling constant in this normalization.  Correspondingly, we refer to
the adjoint field in this normalization as $\phi_2$.  In this
normalization, singularities ({\it e.g.}\/, massless monopoles/dyons,
roots of the baryonic branch) occur on the Coulomb branch of the
theory where a symmetric polynomial of the eigenvalues of the adjoint
field $\phi$ takes special values proportional to the dynamical
scale $\Lambda_2$ in $N=2$ normalization:
\begin{equation}
  \phi_2^k = c_k \Lambda_2^k \equiv c_k
     \left(M e^{-8\pi^2/g_2^2(2N_c-N_f)}\right)^k
  \label{eq:singularities}
\end{equation}
where we used $b_0 = -(2N_c - N_f)$, and $c_k$ are appropriate constants.

One can ask the question whether the locations of such singularities
are RG invariant.  They are indeed RG invariant in an obvious manner
in $N=2$ normalization.  First of all, we never need to change the
normalization of the adjoint field $\phi$ because the normalization
always stays $1/g_2^2$ automatically due to the $N=2$ supersymmetry.
Therefore, the left-hand side of Eq.~(\ref{eq:singularities}) is RG
invariant.  Moreover, there is no $Z_\phi$ contribution to $1/g_2^2$
when one changes the cutoff from $M$ to $M'$.  Second, there is no
wavefunction renormalization for the hypermultiplets \cite{WLVP,APS}.
Therefore, there is no $Z_Q$, $Z_{\tilde{Q}}$ contribution to
$1/g_2^2$ either.  As a result, $\Lambda$ is RG invariant, 
the right-hand side of Eq.~(\ref{eq:singularities}) is also RG
invariant, and hence Eq.~(\ref{eq:singularities}) remains the same
under the change of the cutoff and bare parameters trivially.

If one employs $N=1$ language, the analysis is far less obvious.
First of all, the holomorphic gauge coupling $g_1^2$ in $N=1$ language
differs from $g_2^2$ because one scales the adjoint field to make it
canonically normalized $\phi_1 = g_2^{-1} \phi_2$, and the anomalous
Jacobian \cite{KS} gives
\begin{equation}
  \frac{8\pi^2}{g_1^2} = \frac{8\pi^2}{g_2^2} + N_c \ln g_2^2 .
\end{equation}
The dynamical scale $\Lambda_1= M e^{-8\pi^2/g_1^2(2N_c-N_f)}$ in
$N=1$ normalization is then related to that in $N=2$ normalization by
\begin{equation}
  \Lambda_1 = \Lambda_2 (g_2^2)^{-N_c/(2N_c-N_f)} .
\end{equation}
When one changes the cutoff from $M$ to $M'$, the adjoint field
$\phi_1$ receives a wave function renormalization
\begin{equation}
  Z_\phi(M', M) = \frac{g_2^2}{g'^2_2} .
  \label{eq:Z_phi}
\end{equation}
such that the superpotential coupling $\int d^2\theta \sqrt{2} g_2
\tilde{Q} \phi_1 Q$ is always related to the gauge coupling constant
as required by $N=2$ supersymmetry.

The locations of singularities are now written as
\begin{equation}
  \phi_1^k = c_k g_2^{-k} \left(\Lambda_1
    (g_2^2)^{N_c/(2N_c-N_f)}\right)^k .
  \label{eq:singularities2}
\end{equation}
Now this form is RG invariant under the same analysis as we did
before.  Under the change of the cutoff and bare parameters, the
left-hand side is replaced by
\begin{equation}
  \phi_1'^k = Z_\phi(M', M)^{k/2} \phi_1^k ,
\end{equation}
while the right-hand side by
\begin{eqnarray}
\lefteqn{
  g'^{-k}_2 \left(\Lambda'_1
    (g'^2_2)^{N_c/(2N_c-N_f)}\right)^k } \nonumber \\
& &  =  g'^{-k}_2 \left(\Lambda_1 Z_\phi^{N_c/(2N_c-N_f)}
    Z_Q^{N_f/2(2N_c-N_f)} Z_{\tilde{Q}}^{N_f/2(2N_c-N_f)} 
    (g'^2_2)^{N_c/(2N_c-N_f)}\right)^k . \nonumber \\
\end{eqnarray}
The Eq.~(\ref{eq:singularities2}) remains invariant with $Z_Q =
Z_{\tilde{Q}} = 1$ and Eq.~(\ref{eq:Z_phi}).

\section{Inverted Hierarchy}
\setcounter{equation}{0}
\setcounter{footnote}{0}

In this section we apply our understanding of the RG in supersymmetric
gauge theories to a model of dynamical supersymmetry breaking, where
such an understanding is necessary to resolve puzzles about the
correct vacuum structure of the theory.  The theories we consider are
vector-like SP($N$) models with $N+1$ flavors studied by Izawa, Yanagida
\cite{IY} and by Intriligator, Thomas \cite{IT}.  The question we ask
is whether the so-called inverted hierarchy mechanism \cite{Witten}
operates in these models.  The inverted hierarchy refers to the
situation where dynamics forces the expectation value of a scalar
field to be exponentially large compared to the energy scale of the
potential.  This occurs in O'Rafeartaigh type models of supersymmetry
breaking where the scalar field is a classical flat direction.  The
effective potential is modified by perturbative corrections both from
the gauge coupling and Yukawa coupling.  The potential minimum arises
where the two corrections balance against each other.

The particle content of the SP($N$) models consists of 2(N+1) SP($N$)
fundamentals $Q^i$ and singlets $S_{ij}$. The superpotential is given
by
\begin{equation}
W = \frac{1}{2} \lambda S_{ij} Q^{i} Q^{j}.
\end{equation}
The equation of motion for $S_{ij}$ demands that $Q^{i} Q^{j}=0$,
which is in conflict with the quantum modified constraint $\mbox{Pf}(Q^i Q^j) =
\Lambda^{2(N+1)}$, and supersymmetry is broken. For non-zero $S_{ij}$,
the flavors become massive and can be integrated out of theory. The
resulting low-energy theory is pure SP($N$) with a dynamical scale
depending on $S_{ij}$. This theory exhibits gaugino condensation and
generates an effective superpotential for $S_{ij}$
\begin{equation}
  W_{\!\it eff}(S_{ij}) = (\mbox{Pf}\lambda S_{ij})^{1/2(N+1)}
  \Lambda^2.
  \label{eq:Weff}
\end{equation}
If we expand $S_{ij}$ around $S_{ij} = \sigma J_{ij}/\sqrt{N+1}$ where $J_{ij}$
is the symplectic matrix, all components of $S_{ij}$ other than
$\sigma$ become massive. The effective potential for $\sigma$ is then
\begin{equation}
W_{\!\it eff}(\sigma) = \lambda \sigma \Lambda^2.
\end{equation}    
The $\sigma$ equation of motion shows that supersymmetry is
broken, while the tree-level potential for $\sigma$ is 
\begin{equation}
  V_{\it tree} (\sigma) = |\lambda \Lambda^2|^2,
  \label{eq:Vtree}
\end{equation}
and the vev of $\sigma$ is undetermined at this level.
Since supersymmetry is broken, we expect that some
nontrivial potential will be generated for $\sigma$ at higher orders
in perturbation theory. In \cite{Yuri}, it was argued that the
potential $V(\sigma)$ is ``RG improved" as $V(\sigma)=|\lambda
\Lambda^2|^2 \rightarrow |\lambda(\sigma) \Lambda^2|^2$, where
$\lambda(\sigma)$ is the running value of $\lambda$, which receives
contributions from both the $S_{ij}$ and $Q^{i}$ wavefunction
renormalizations $\lambda(\sigma) = Z_S^{-1/2}(\sigma)
Z_Q(\sigma)^{-1} \lambda$.  If this conclusion is correct, it is
possible to realize the inverted hierarchy mechanism in this model:
$\sigma$ could develop a stable vev much larger than $\Lambda$, since
the $Z_S$ factors depend on $\lambda$ and tend to make the potential
rise for large $\sigma$, whereas the $Z_Q$ factors depend on
asymptotically free SP($N$) gauge coupling and tend to make the
potential rise for small $\sigma$.

On the other hand, a host of arguments indicate that this conclusion
can not be correct. For instance, the superpotential for $\sigma$ is
exact, and so the potential is only modified by the K\"ahler potential
for $\sigma$, which at one loop only depends on $\lambda$.
Alternately, $F_\sigma = \lambda \Lambda^2$ generates a
non-supersymmetric spectrum for the $Q^i$ but not for the gauge
multiplet. If we simply look at the 1-loop effective potential, the
only contribution comes from the non-supersymmetric $Q^i$ spectrum and
again depends only on $\lambda$, and the potential is monotonically
increasing with $\sigma$.

In order to resolve this puzzle, we must carefully consider how the
potential is ``RG improved". As we will show explicitly in the
remainder of the section, the solution is that not only
$\lambda$ but also $\Lambda$ runs; the correct RG improvement of the
potential is $V(\sigma) = |\lambda(\sigma) \Lambda(\sigma)^2|^2$, and
all the $Z_Q$ dependence cancels in the product $\lambda(\sigma)
\Lambda^2(\sigma)$.

Let us go back to the start and carefully define the problem. With
cutoff $M$, the Lagrangian is given by
\begin{equation}
  {\cal L} =
  \int d^4\theta \left(Q^{i \dagger} e^{2V} Q^{i} + 
    \frac{1}{2} \mbox{Tr} S^{\dagger} S\right) 
   + \int d^2 \theta \left(\frac{1}{4 g^2_h} W^a W^a + 
  \frac{1}{2} \lambda S_{ij} Q^i Q^j\right) + h.c. 
\end{equation}
In the functional integral, we would like to integrate out the $Q^i$
and the gauge multiplet and be left with an effective Lagrangian for
$S_{ij}$:
\begin{equation}
  \exp\left(-\int d^4x {\cal L}_{\it eff}(S)\right) \equiv
  \int DV DQ \mbox{exp}\left(-\int d^4x {\cal L}\right),
\end{equation}
with
\begin{eqnarray} 
  {\cal L}_{\it eff}(S) = \int d^4 \theta 
  \left(\frac{1}{2}\mbox{Tr} S^{\dagger} S
  + \delta K(\lambda S, \lambda^{\dagger} S^{\dagger}) \right)
  + \int d^2 \theta W_{\!\it eff} (\lambda S) + h.c.,
\label{eq:Leff}
\end{eqnarray}  
where both $\delta K$ and $W_{\!\it eff}$ depend further on the gauge
coupling and the cutoff, $(1/g^2_h, M)$, or equivalently on
$(\Lambda, M)$.

Of course, when we integrate out the $Q,V$ multiplets in perturbation
theory, we never generate any effective superpotential.  However, a
superpotential is generated non-perturbatively, and its form is
completely determined by a non-anomalous $R$ symmetry under which
$\lambda S_{ij}$ has charge $+2$, an anomalous U(1) symmetry under
which $\lambda S_{ij}$ has charge $-2$ and $\Lambda$ has charge $+1$,
and the non-anomalous SU($2N+2$) flavor symmetry.  Together, these dictate
(up to an overall constant) the exact effective superpotential
Eq.~(\ref{eq:Weff}).  One finds an effective potential along the
$S_{ij}=\sigma J_{ij}/\sqrt{N+1}$ direction:
\begin{equation}
  V(\sigma;\lambda,\Lambda;M) =
  \frac{|\lambda \Lambda^2|^2}{1 + \left.\left(\partial^2 \delta K/\partial S
        \partial S^{\dagger}\right)\right|_{S=\sigma J}} \ .
  \label{eq:Veff}
\end{equation}

The corrections to the K\"ahler potential $\delta K(\lambda
S,\lambda^{\dagger} S^{\dagger})$ are certainly generated in
perturbation theory.  For instance, at 1-loop there is a contribution
from the loop of $Q$'s, which yields (at the leading-log)
\begin{equation}
  \delta K(\lambda S, \lambda^{\dagger} S^{\dagger}) = 
  - 2 \frac{N}{16 \pi^2} (\lambda \sigma)^{*} (\lambda \sigma)
  \mbox{ln} \frac{|\lambda\sigma|^2}{M^2}
\end{equation}
along the $\sigma$ direction and we find:
\begin{equation}
  V(\sigma;\lambda,\Lambda;M)= |\lambda \Lambda^2|^2
  \left[1 + 2 \frac{N\lambda^2}{16\pi^2}
      \left(\ln\frac{|\lambda\sigma|^2}{M^2} +4 \right) 
  + {\cal O}(\lambda^4) \right] .
\end{equation}

For $\sigma\ll M$, the large logs in the above expression make
perturbation theory unreliable.  However, we can use the same
technique as in Sec.~2 to deal with this problem. We are interested in
$V(\sigma;\lambda,\Lambda; M)$. First, we rescale everything by naive
dimensional analysis
\begin{equation}
  V(\sigma;\lambda,\Lambda;M)= 
    e^{-4 t} V(e^t \sigma; \lambda,e^t\Lambda,;e^t M). 
    \label{eq:scaling}
\end{equation}
Next, we bring the cutoff back from $e^t M$ to $M$ by appropriately
changing the couplings.  Since the form of the tree potential is known 
with canonical normalization for the superfields $Q$, $S$, we would
like to keep them canonical.  Hence, as we have argued in Sec.~2, not only 
$\lambda$ but also $\Lambda$ must be changed:
\begin{equation}
  V(e^t \sigma; \lambda, e^t \Lambda; e^t M) = 
  V(e^t \sigma Z_S^{1/2}(t); \lambda Z_Q^{-1}(t) Z_S^{-1/2}(t), 
    e^t \Lambda Z_Q(t)^{1/2}; M),
\end{equation}
where $Z_Q(t)\equiv Z_Q(M, e^t M)$, $Z_S(t) \equiv Z_S(M, e^t M)$.
However, we can choose $e^t |\sigma| \sim M$, then the logarithms in
the perturbative expansion of the RHS are small and the tree value of
the RHS $V_{\it tree} (\sigma; \lambda, \Lambda; M) = |\lambda
\Lambda^2|^2$ is an excellent approximation.  Doing this and combining
with Eq.~(\ref{eq:scaling}), we find
\begin{eqnarray}
  V(\sigma;\lambda,\Lambda;M) &\simeq& 
  \left|(\lambda Z_Q^{-1}(t)Z_S^{-1/2}(t))(\Lambda Z_Q^{1/2}(t))^2
  \right|^2_{t \sim \ln (\lambda \sigma/M)} \nonumber \\
  &=& Z_S^{-1}(\lambda\sigma, M) |\lambda \Lambda^2|^2.
\end{eqnarray}
As promised, when the RG improvement is done consistently, the $Z_Q$
dependence drops out and we are left with a monotonically rising
potential (from the $Z_S$ factor) which does not realize the inverted
hierarchy.  

One can also formally check that the effective potential $V(\sigma;
\lambda,\Lambda; M)$, or equivalently the effective Lagrangian
Eq.~(\ref{eq:Leff}), is RG invariant, {\it i.e.}\/ independent of the
choice of the cutoff $M$ as long as one changes the bare couplings
appropriately.  First of all, the effective superpotential
Eq.~(\ref{eq:Weff}) is invariant by itself, because
\begin{eqnarray}
\lefteqn{
  \lambda' (\mbox{Pf} S'_{ij})^{1/2(N+1)} \Lambda'^2 } \nonumber \\
  &=& \! \! \! \lambda Z_Q^{-1}(M', M) Z_S^{-1/2}(M', M)
  (\mbox{Pf} S_{ij})^{1/2(N+1)} Z_S^{1/2}(M', M)
  (\Lambda Z_Q^{1/2}(M',M))^2 \nonumber \\
  &=& \! \! \!\lambda (\mbox{Pf} S_{ij})^{1/2(N+1)} \Lambda^2.
\end{eqnarray}
The K\"ahler potential along the $\sigma$ direction
\begin{equation}
  \int d^4 \theta Z_\sigma (\lambda \sigma, M) \sigma^* \sigma
\end{equation}
is also RG invariant which can be seen as follows.  First, the
$Z_\sigma$ factor must depend on the renormalized effective mass of
$Q$, $m_Q \equiv \lambda \sigma Z_Q^{-1}(m_Q, M)$ because it is
generated from integrating out the massive $Q$ field.  This
combination can be easily seen to be RG invariant.  Second, it is
multiplicative, $Z_\sigma(m_{Q}, M) =Z_\sigma(m_{Q},
M') Z_\sigma(M', M)$.  With the definition $\sigma' = Z_\sigma^{1/2}
(M', M) \sigma$, the K\"ahler potential is RG invariant.  Since the
change of the field variable $\sigma$ to $\sigma'$ is given by a 
field-independent constant $Z_\sigma(M, M')$, the
auxiliary equation for the $F_\sigma$ changes only by an overall
factor $Z_\sigma(M, M')$.  On the other hand the quadratic term of 
$F_{\sigma}$ also changes the the same factor and hence we conclude that 
the effective potential (\ref{eq:Veff}) is RG invariant.

We have demonstrated that the inverted hierarchy mechanism does not
work in these models, contrary to the naive argument of RG improvement
\cite{Yuri}.  In order to achieve the inverted hierarchy mechanism as
favored from the model building point of view, one needs to make the
flat direction fields $S_{ij}$ gauge non-singlet, as was recently done in
\cite{HM,DDGR}.

\section{Conclusion}
\setcounter{equation}{0} 
\setcounter{footnote}{0}

In this paper, we studied the renormalization group invariance of the 
exact results in supersymmetric gauge theories.  We first clarified 
the notion of Wilsonian renormalization group (RG) invariance in 
supersymmetric gauge theories.  It is a non-trivial statement that the 
low-energy physics can be kept fixed when one changes the ultraviolet 
cutoff with appropriate changes in the bare coupling constants in the 
Lagrangian.  We derived the formula for the changes of bare couplings 
using two methods: one using strictly Wilsonian actions and 
holomorphic gauge coupling, the other using the perturbative NSVZ 
$\beta$-function.  We used canonical normalization for the chiral 
superfields because it allows the most straightforward application of 
the renormalization group.  We find that the scale $\Lambda$ is {\it 
not}\/ RG invariant.  We then compared our formula to known exact 
results and showed that they actually require the changes in $\Lambda$ 
we have derived.

Finally, we applied our result to models of supersymmetry breaking due
to quantum modified constraints, namely SP($N$) models with $N+1$
flavors.  These models have a classically flat direction, and the crucial
question is in what way the flat direction is lifted.  The RG
invariance allowed us to determine the effective potential along the
classical flat direction.  A naive application of RG improvement of
the potential would tell us that the potential along the 
flat direction is modified
perturbatively both by the SP($N$) gauge interaction and the
superpotential interaction, and hence that the flat direction may develop
an expectation value exponentially larger than the supersymmetry
breaking scale (inverted hierarchy mechanism).  However, a careful
application of our method demonstrates that the inverted hierarchy
mechanism does not occur in these models.

\section*{Acknowledgments}
This work was supported in part by the U.S. Department of 
Energy under Contract DE-AC03-76SF00098, in part by the National 
Science Foundation under grant PHY-95-14797.  NAH was also supported
by NERSC, and HM by the Alfred P. Sloan Foundation.


\begin{thebibliography}{99}
  
\bibitem{IS} K. Intriligator and N. Seiberg, lectures presented at
  TASI 95, Boulder, CO, Jun 4-26, 1995 and Trieste Spring School,
  Trieste, Italy, Mar 27 - Apr 15, 1995 and Trieste Summer School,
  Trieste, Italy, Jun 12 - Jul 23, 1995 and Carge\'se Summer School,
  Carg\'ese, France, Jul 11-24, 1995, hep-th/9509066.  {\sl Nucl. Phys.
    Proc. Suppl.}\/ {\bf 45BC} 1-28 (1996)
  
\bibitem{SV} M. A. Shifman and A. I. Vainshtein, {\sl Nucl. Phys.}\/
  {\bf B277}, 456 (1986).
  
\bibitem{DS} M. Dine and Y. Shirman, {\sl Phys. Rev.}\/ {\bf D50},
  5389 (1994).

\bibitem{FP} D. Finnell and P. Pouliot, {\sl Nucl. Phys.}\/ {\bf
    B453}, 225 (1995).

\bibitem{KS} K. Konishi and K. Shizuya, {\sl Nuovo Cim.}\/ {\bf 90A}
  111 (1985).

\bibitem{AM2} N. Arkani-Hamed and H. Murayama, UCB-PTH/97/22, in
  preparation. 

  
\bibitem{JJN}I. Jack, D. R. T. Jones, C. G. North, {\sl Nucl. Phys.}\/
  {\bf B486} 479 (1997).
  
\bibitem{NSVZ} V. Novikov et al, {\sl Nucl. Phys.}\/ {\bf B229} 381
  (1983); {\sl Phys. Lett.}\/ {\bf B166} 329 (1986).
  
\bibitem{SP(N)} K. Intriligator and P. Pouliot, {\sl Phys. Lett.}\/
  {\bf B353}, 471 (1995).

\bibitem{PT} E. Poppitz and S. P. Trivedi, {\sl Phys. Lett.}\/ {\bf
    B365}, 125 (1996).

\bibitem{WLVP} B. de Wit, P. G. Lauwers, and A. Van Proeyen, {\sl
    Nucl. Phys.}\/ {\bf B255} 569 (1985).

\bibitem{APS}  P. C. Argyres, M. R. Plesser, and N. Seiberg, {\sl
    Nucl. Phys.}\/ {\bf B471}, 159 (1996).

\bibitem{IY} K.-I. Izawa and T. Yanagida, hep-th/9602180, {\sl Prog.  
Theor. Phys.}\/ {\bf 95}, 829 (1996).

\bibitem{IT} K. Intriligator and S. Thomas, hep-th/9603158, {\sl 
Nucl. Phys.}\/ {\bf B473}, 121 (1996).

\bibitem{Witten} E.~Witten, {\sl Phys. Lett.}\/ {\bf 105B}, 267
  (1981).

\bibitem{Yuri} Y.~Shirman, hep-th/9608147, {\sl Phys. Lett.}\/ 
{\bf B389}, 287 (1996). 

\bibitem{HM}
H. Murayama, LBL-40286, UCB-PTH/97/20, hep-ph/9705271.

\bibitem{DDGR}
S. Dimopoulos, G. Dvali, R. Rattazzi, and G.F. Giudice,
CERN-TH-97-098, hep-ph/9705307.  



\end{thebibliography}
\end{document}